\documentclass[twoside]{photon2007}
\usepackage[latin1]{inputenc}
\usepackage{graphicx,epsfig,color}
\usepackage{wrapfig,rotating}
\usepackage{amssymb,amsmath,array}

\newcommand{\be}{\begin{equation}}
\newcommand{\ee}{\end{equation}}
\newcommand{\bc}{\begin{center}}
\newcommand{\ec}{\end{center}}
\begin{document}
\title{
%%%%   Paper title goes here  %%%%%%%%%%%%%%
Sense and Nonsense on Parton Distribution functions of the Photon} 
%% 
%***********************************************************************
% AUTHORS INFORMATION AREA
%***********************************************************************
\author{Ji\v{r}\'{\i} Ch\'{y}la
% Optional short acknowledgment: remove next line if non-needed
%\thanks{This is an optional funding source acknowledgment.}
% DO NOT MODIFY THE FOLLOWING '\vspace' ARGUMENT
\vspace{.3cm}\\
% Addresses and institutions (remove "1- " in case of a single institution)
Institute of Physics, Academy of Sciences of the Czech Republic\\
Na Slovance 2, 18221 Prague 8, Czech Republic 
}
%%***********************************************************************
% END OF AUTHORS INFORMATION AREA
%***********************************************************************

\maketitle

\begin{abstract}
The organization of finite order QCD approximations to $F_2^{\gamma}(x,Q^2)$
based on the separation of pure QED contribution
% to $F_2^{\gamma}(x,Q^2)$
from those of genuine QCD nature is discussed. 
\end{abstract}
\section{Introduction}
Using the recently completed calculation of order $\alpha_s^3$ parton-parton
and order $\alpha\alpha_s^2$ photon-parton splitting
functions \cite{MVVph1,VogtWarsaw} the standard NLO and NNLO approximations for the photon structure function $F_2^{\gamma}(x,Q^2)$ exhibit very small
difference when the latter is evaluated in the 
$\overline{\mathrm{MS}}$ factorization scheme (FS). This stands in contrast to 
the large difference in this FS, noted in \cite{GRV92}, between the LO and NLO approximations. In order to cure this, and related problems, the so called DIS$_\gamma$ FS has been proposed in \cite{GRV92b}.

In this talk I will argue that this large difference between the standard LO and 
NLO approximations to $F_2^{\gamma}$ in the $\overline{\mathrm{MS}}$ FS disappears provided these approximations are defined with respect to genuine QCD contribution. I will furthermore argue that the way DIS$_\gamma$ FS is introduced violates the factorization scheme invariance of $F_2^{\gamma}$ and is thus theoretically flawed. 
For lack of space only brief outline of the arguments can be given in this written version, for details see \cite{ja1,ja2}. 

The source of the problem in the standard treatment of $F_2^{\gamma}$ can 
be traced back to the interpretation of the behaviour of PDF of the photon in perturbative QCD. 
In \cite{ja1} I have discussed this point at length and argued 
that PDF of the photon behave like $\alpha$, rather than $\alpha/\alpha_s$
as usually claimed. To see the point, let us
recall basic facts concerning the theoretical analysis of 
$F_2^{\gamma}$. For simplicity we shall restrict ourselves to the nonsinglet
channel, where 
\begin{equation}
\frac{F_{2,{\mathrm NS}}^{\gamma}(Q^2)}{x}= q_{\mathrm{NS}}(M)\otimes
C_q(Q/M)+\delta_{\mathrm{NS}}C_{\gamma}
\label{NSpart} \nonumber
\end{equation}
with $q_{\mathrm{NS}}$ satisfying the evolution equation 
\begin{equation}
\frac{{\mathrm d}q_{\mathrm {NS}}(x,M)}{{\mathrm d}\ln M^2} 
= \delta_{\mathrm{NS}}k_q
+P_{\mathrm {NS}}\otimes q_{\mathrm{NS}}
\label{NSevolution}
\end{equation}
and $\delta_{\mathrm{NS}}=6n_f\left(\langle e^4\rangle-\langle
e^2\rangle ^2\right)$. The splitting functions $P_{{\mathrm{NS}}}$ and
$k_q$ are given as power expansions in $\alpha_s(M)$:
\begin{eqnarray}
k_q &\!\!\!\!\! = &\!\!\!\!\! \frac{\alpha}{2\pi}\left[k^{(0)}_q(x)+
\frac{\alpha_s(M)}{2\pi}k_i^{(1)}(x)+\cdots\right],\nonumber
\label{splitquark} \\
P_{{\mathrm {NS}}} &\!\!\!\!\! =&\!\!\!\!\! 
~~~~~~~~~~~~~~~~~~~\frac{\alpha_s(M)}{2\pi}P^{(0)}_{{\mathrm{NS}}}(x)+\cdots.
\nonumber
\label{splitpij}
\end{eqnarray}
The leading order splitting functions
$k_q^{(0)}(x)=x^2+(1-x)^2$ and $P^{(0)}_{{\mathrm{NS}}}(x)$ are {\em
unique}, whereas all higher order ones depend on the choice of
the {\em factorization scheme} (FS). The coefficient functions
$C_q,C_{\gamma}$ admit perturbative expansions of the form
\begin{eqnarray}
C_q(x,Q/M)&\!\!\!\!\!  = &\!\!\!\!\! \delta(1-x)+
\frac{\alpha_s(\mu)}{2\pi}C^{(1)}_q+\cdots,\nonumber
\label{cq} \\
C_{\gamma}(x,Q/M)& \!\!\!\!\! = &\!\!\!\!\!\frac{\alpha}{2\pi}\left[
C_{\gamma}^{(0)}+
\frac{\alpha_s(\mu)}{2\pi}C_{\gamma}^{(1)}+\cdots\right],\nonumber
\label{cg}
\end{eqnarray}
where coefficient function $C_{\gamma}^{(0)}$
\begin{equation}
C_{\gamma}^{(0)}=k_q^{(0)}(x)\ln\frac{Q^2(1-x)}{M^2x}+8x(1-x)-1\nonumber
\label{C0}
\end{equation}
similarly as $k_q^{(0)}$ is of the pure QED origin. To simplify the formulae I will in the following 
drop the subscript ``NS'' and set $\delta_{\mathrm{NS}}=1$ everywhere.

\section{PDF of the photon}
\begin{figure}[h]
\includegraphics[width=1.0\columnwidth]{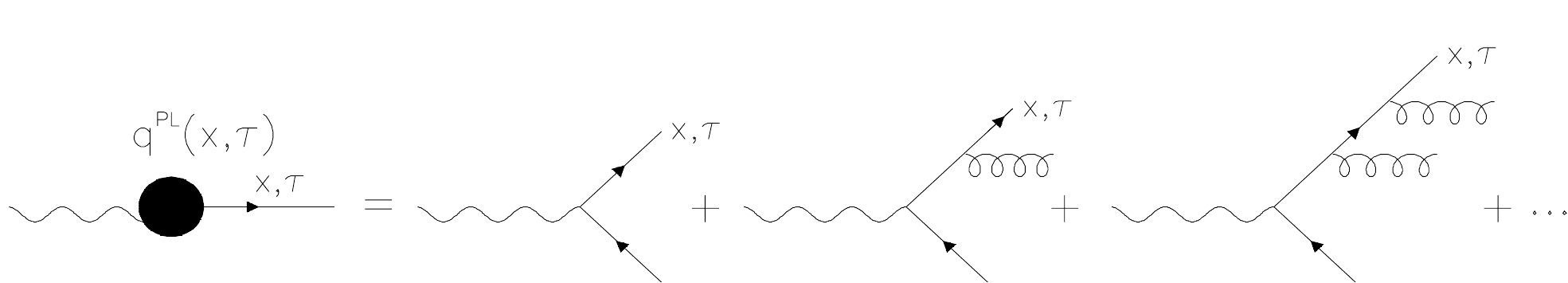}
\caption{Diagrams defining the pointlike part of $q_{\mathrm{NS}}$.}
\label{figpl}
\end{figure}

The general solution of the evolution equation (\ref{NSevolution}) can be 
written as the sum of a particular solution of the full inhomogeneous equation 
and the general solution of the corresponding homogeneous one, called
{\em hadronic}. The subset of the solutions of (\ref{NSevolution}) resulting
from the resummation of the contributions of diagrams in Fig. \ref{figpl} 
and vanishing at some definite $M_0$, are called {\em pointlike} solutions. 
In the LLA, i.e. taking only the first terms $k^{(0)}$ and $P^{(0)}$ on the r.h.s. of (\ref{NSevolution}), it has in momentum space the closed form
\begin{eqnarray}
\lefteqn{ q_{\mathrm{PL}}(n,M_0,M)=\frac{\displaystyle 4\pi a(n)}{\alpha_s(M)}}
 \nonumber\\&&
 \left[1-\left(\frac{\alpha_s(M)}{\alpha_s(M_0)}\right)^
{1-2P^{(0)}(n)/\beta_0}\right].
\label{generalpointlike}
\end{eqnarray}
As the principal difference between PDF of the photon and hadrons result from 
the contribution to the former of these pointlike solutions, only they will be discussed further and the specification ``pointlike'' will be dropped. It must, however, be kept 
in mind that the separation of PDF of the photon into hadronic and pointlike 
parts is not unique. In our simplified case it depends on the choice of $M_0$.
The smaller $M_0$, the more important is the pointlike part relative to the
hadronic one and vice versa.

The fact that $\alpha_s(M)$ appears in the denominator of (\ref{generalpointlike}) is usually, and unfortunately, interpreted literally, i.e. as implying that the pure QED contribution $C_{\gamma}^{(0)}$ is of ``higher order'' than quark distribution
function $q$. This, in turn, implies that the former is not included in the standard ``LO'' approximation to $F_2^{\gamma}(x,Q^2)$
\begin{equation}
F_{2,{\mathrm NS}}^{\gamma,{\mathrm{LO}}}(x,Q^2)= x q_{\mathrm{NS}}(x,M)
\label{NSLO}
\end{equation}
which also ``behaves'' as $\alpha/\alpha_s$. This is obvious nonsense as if we switch off QCD we must get pure QED contribution. Indeed, keeping $M$ and $M_0$ fixed and switching QCD off by
sending $\Lambda\rightarrow 0$ the expression (\ref{generalpointlike})
approaches 
\begin{equation}
q(M,M_0)\rightarrow q_{\mathrm{QED}}\equiv
\frac{\alpha}{2\pi} k^{(0)}
\ln\frac{M^2}{M_0^2}\, ,
\label{QEDlimit}
\end{equation}
corresponding to the first, pure QED, diagram in Fig. \ref{figpl}.
To see what is wrong with the claim that $q\propto \alpha/\alpha_s$ consider the pure QED evolution equation
\begin{equation}
\frac{{\mathrm d}q_{\mathrm {NS}}(x,M)}{{\mathrm d}\ln M^2} 
=\frac{\alpha}{2\pi}k^{(0)}_q(x)
\label{QEDNSevolution}
\end{equation}
There is no trace of QCD, but we can formally rewrite it in terms of the derivative with respect to QCD couplant $\alpha_s$ (taking for simplicity
$\beta$-function to the lowest order)
\begin{equation}
\frac{\mathrm{d}q(n,Q)}{\mathrm{d}\alpha_s}=
-\frac{4\pi}{\beta_0}\frac{\alpha}{2\pi}\frac{k^{(0)}(n)}{\alpha_s^2}.
\label{NSequiv}
\end{equation}
This equation can be trivially solved to get
\begin{equation}
q(n,Q)=
\frac{\alpha}{2\pi}\frac{4\pi}{\beta_0}\frac{k^{(0)}(n)}{\alpha_s(Q)}+A,
\label{sol}
\end{equation}
where $A$ stands for arbitrary integration constant specifying the boundary
condition on the solution of (\ref{NSequiv}), which can be chosen as $A=0$.
The above expression (\ref{sol}) might, but should not, mislead us to the usual claim that $q\propto \alpha/\alpha_s$, because we know that we are still within pure QED! Indeed, if we evaluate the diiference at two scales and insert the explicit expression for $\alpha_s(Q)$ we get back the starting, purely QED expression
\begin{equation}
q(n,Q_1)-q(n,Q_2)=
\frac{\alpha}{2\pi}k^{(0)}(n)\ln\frac{Q_1^2}{Q_2^2}
\label{man3}
\end{equation}
There is another argument demonstrating that PDF of the photon 
behave as ${\cal O}(\alpha)$. Consider the Mellin moments 
\footnote{To save the space the dependence on the Mellin moment
variable $n$ will henceforth be suppressed.}
of $F^{\gamma}(x)\equiv F_{\mathrm{NS}}^{\gamma}(x,Q)/x$
\begin{equation}
F^{\gamma}(Q)=q(M) C_q(Q/M)+C_{\gamma}(Q/M)
\label{FNS}
\end{equation}
where $q(M)$ vanishes at $M_0$: $q(M_0)=0$. 
As $F^{\gamma}(Q)$ is independent of the factorization scale
$M$, we can take any $M$ to evaluate (\ref{FNS}), for instance
$M_0$. However, for $M=M_0$ the first term in (\ref{FNS}) vanishes
and we get
\begin{eqnarray}
&&\!\!\!\!\!\!\!\!\!\!\!\!\!\frac{2\pi}{\alpha}F^{\gamma}(Q)=C_{\gamma}^{(0)}\left(\frac{Q}{M_0}\right)+
\frac{\alpha_s(\mu)}{2\pi}C_{\gamma}^{(1)}\left(\frac{Q}{M_0}\right)\nonumber\\&&+
\left(\frac{\alpha_s(\mu)}{2\pi}\right)^2C_{\gamma}^{(2)}\left(\frac{Q}{M_0},\frac{Q}{\mu}\right)
+\cdots 
\label{cgammaonly}
\end{eqnarray}
i.e. manifestly the expansion in powers of
$\alpha_s(\mu)$, which starts with ${\cal O}(\alpha)$ pure QED
contribution $(\alpha/2\pi)C_{\gamma}^{(0)}$ and includes standard
QCD corrections of orders $\alpha_s^k,k\ge 1$, which vanish when QCD is
switched off, and no trace of the supposed ``$\alpha/\alpha_s$'' behaviour. 

\section{From semantics to substance}
The claim that $q\propto \alpha/\alpha_s$ stems in part from inappropriate
terminology. Although a matter of convention, it is wise to
define the terms ``leading'', ``next--to--leading'' etc. 
in a way which guarantees that they have the same meaning in different
processes. For the case of the ratio
\begin{eqnarray}
&& \!\!\!\!\!\!R_{\mathrm{e}^+\mathrm{e}^-}(Q)\equiv
\frac{\sigma(\mathrm{e}^+\mathrm{e}^-\rightarrow
\mathrm{hadrons})}{\sigma(\mathrm{e}^+\mathrm{e}^-\rightarrow
\mu^+\mu^-)}=\nonumber \\&&
\left(3\sum_{i=1}^{n_f}e_i^2\right)\left(1+r(Q)\right)
\label{Rlarge}
\end{eqnarray}
it is general practice to apply the terms ``leading'' and ``next--to--leading'' 
to genuine QCD effects as described by $r(Q)$, i.e. subtracting from (\ref{Rlarge}) the pure QED contribution. Unfortunately, this practice is ignored in most analyses
of $F^{\gamma}$ \cite{GRV92,GRV92b}.

In \cite{ja1} I have proposed the definition
of QCD approximations to $F^{\gamma}$ which follows closely QCD analysis of quantities like (\ref{Rlarge}).
It starts with writing the quark distribution
function $q(M)$ as the sum of the purely QED contribution (\ref{QEDlimit}) 
%\begin{equation}
%q_{\mathrm{QED}}(M)\equiv\frac{\alpha}{2\pi}k^{(0)}
%\ln\frac{M^2}{M_0^2}
%\label{qQED}
%\end{equation}
and the QCD corrections, which start at order $\alpha_s$ and are generated
by the terms in splitting and coefficient functions proportional to positive
powers of $\alpha_s$. This implies significant difference in the definition
of fixed order approximations compared to those in the standard aproach. Whereas the standard LO approximation to $F_2^{\gamma}$ includes only the lowest order splitting functions $k^{(0)}_q$ and $P^{(0)}_{\mathrm{NS}}$, we include four more terms. $C^{(0)}_{\gamma}$, which is of pure QED origin and is closely related to $k^{(0)}$, as well as three terms appearing at order $\alpha_s$: $k^{(1)}_q$, $C^{(1)}_{\gamma}$ 
and $C_q^{(1)}$. At the NLO the difference is even more pronounced (see table 1 of \cite{ja2}). The numerical importance of the individual contributions of these additional terms as well as the overall difference between the standard and our definition of the LO approaximation to  $F_2^{\gamma}$ has
been discussed by J. Hejbal \cite{hejbal}. The latter is 
phenomenologically relevant and comparable to errors of existing data.    

\section{What is wrong with the DIS$_{\gamma}$ factorization scheme?}
In \cite{ja2} I discussed how the freedom in the definition of 
quark distribution functions of the photon is related to the non-universality of the coefficient functions $C_q^{(j)}, j\ge 1$. The QED contribution to $F^{\gamma}$ is of order $\alpha_s^0=1$, but since in the standard approach
the quark distribution function is assigned to the order $1/\alpha_s$, the QED photonic coefficient function $C_{\gamma}^{(0)}$ appears as ``NLO'' contribution 
and is therefore treated in a similar way as the lowest order QCD
coefficient function $C_q^{(1)}$. In \cite{GRV92} the authors introduced
the so called DIS$_{\gamma}$ factorization scheme by absorbing in the redefined quark distribution function of the photon the pure QED term
$C_{\gamma}^{(0)}$ according to (see eq. (5) of \cite{GRV92b})
\be
\overline{q}(M,M_0)\equiv q(M,M_0)+\frac{\alpha}{2\pi}C_{\gamma}^{(0)}(1).
\label{subst}
\ee
and imposing {\em ``the same boundary conditions for the pointlike LO
and HO distributions''}
\be
q^{\gamma}_{\mathrm{PL}}(x,Q^2_0)=
\overline{q}^{\gamma}_{\mathrm{PL}}(x,Q^2_0)=
G^{\gamma}_{\mathrm{PL}}(x,Q^2_0)=0
\label{GRVcit}
\ee
as in the original FS \cite{GRV92}. The redefinition (\ref{subst}) is legitimate
procedure but is in conflict with the assumption (\ref{GRVcit}). To see why remember
that (\ref{subst}) involves QED quantity $C_{\gamma}^{(0)}$, and as
the notion of quark distribution function inside the photon is well-defined
also in pure QED, the above procedure must make sense even in absence of
QCD effects. Moreover, as the QCD contribution depends on the numerical value of $\alpha_s$ it cannot cure any problem of the pure QED part.

In pure QED the contribution to $F^{\gamma}$
coming from the box diagram regularized by explicit quark mass $m_q$ reads
\begin{eqnarray}
\lefteqn{F^{\gamma}_{\mathrm{QED}}(Q)=
\frac{\alpha}{2\pi}C_{\gamma}^{(0)}(Q/m_q)= }
 \nonumber\\&&
\frac{\alpha}{2\pi}\left[k^{(0)}\ln\frac{Q^2}{m_q^2}+
C_{\gamma}^{(0)}(1)\right].
\label{pureQED}
\end{eqnarray}
Introducing an arbitrary scale $M$, we can split it into
quark distribution function
\be
q_{\mathrm{QED}}(M)\equiv \frac{\alpha}{2\pi}
k^{(0)} \ln\frac{M^2}{m_q^2}
\label{qQED2}
\ee
and $C_{\gamma}^{(0)}(Q/M)$ 
\be
F^{\gamma}_{\mathrm{QED}}(Q)=q_{\mathrm{QED}}(M)+
\frac{\alpha}{2\pi}C_{\gamma}^{(0)}(Q/M).
\label{FQED}
\ee
We can redefine $q_{\mathrm{QED}}(M)$ by adding to it an arbitrary function 
$f(n)$ (the DIS$_{\gamma}$ FS of \cite{GRV92} corresponds
to $f=C_{\gamma}^{(0)}(1)$) according to
\be
q_{\mathrm{QED},f}(M)\equiv q_{\mathrm{QED}}(M)+\frac{\alpha}{2\pi}f.
\label{qQED'}
\ee
In order to keep the sum
\be
F^{\gamma}_{\mathrm{QED}}(Q)=q_{\mathrm{QED},f}(M)+
\frac{\alpha}{2\pi}C_{\gamma,f}^{(0)}(Q/M)\nonumber
\label{FQEDf}
\ee
independent of $f$ we must change $C_{\gamma}^{(0)}$ accordingly
\be
C_{\gamma,f}^{(0)}(Q/M)\equiv C_{\gamma}^{(0)}(Q/M)-f.
\label{CQED'}
\ee
We can write down the evolution equation
\be
\frac{{\mathrm d}q_{\mathrm {QED},f}(M)}{{\mathrm d}\ln M^2}=
\frac{\alpha}{2\pi} k^{(0)},
\label{QEDevol}
\ee
which is the same in all $f$-schemes but we are not allowed to 
use the same boundary condition for all $q_{\mathrm{QED},f}(M)$. 
If we do that and impose the boundary condition $q_{\mathrm{QED},f}(M=m_q)=0$ for all $f$ we get
\be
F^{\gamma}(Q)=\frac{\alpha}{2\pi}\left(C_{\gamma}^{(0)}(Q/m_q)-f\right)
\label{fscheme}
\ee
which depends on the choice of $f$ and only for $f=0$ coincides with  (\ref{pureQED}). 

\section{Acknowledgments}

This work has been supported by the the project AV0-Z10100502 of the
Academy of Sciences of the Czech Republic and project LC527 of the Ministry
of Education of the Czech Republic.

% ****************************************************************************
% BIBLIOGRAPHY AREA
% ****************************************************************************

\begin{footnotesize}
% IF YOU DO NOT USE BIBTEX, USE THE FOLLOWING SAMPLE SCHEME FOR THE REFERENCES
% ----------------------------------------------------------------------------

% ----------------------------------------------------------------------------

% IF YOU USE BIBTEX,
% - DELETE THE TEXT BETWEEN THE TWO ABOVE DASHED LINES
% - UNCOMMENT THE NEXT TWO LINES AND REPLACE 'Name_Of_Your_BibFile'

%\bibliographystyle{unsrt}
%\bibliography{Name_Of_Your_BibFile}
% example of Name_Of_Your_BibFile.bib
% @Article{Turcato:2006ch,
%      author    = "Turcato, M.",
%  collaboration = "ZEUS and H1",
%      title     = "Lepton flavour violation and charmonium physics at HERA",
%      journal   = "Nucl. Phys. Proc. Suppl.",
%      volume    = "162",
%      year      = "2006", 
%      pages     = "283-287",
%      SLACcitation  = "%%CITATION = NUPHZ,162,283;%%"
% }
% 
% @Unpublished{Gogitidze:2007du,
%      author    = "Gogitidze, N.",
%  collaboration = "H1", 
%      title     = "Prompt photons and particle momentum distributions at
%                   HERA", 
%      year      = "2007",
%      note    = "hep-ex/0701033",
%      SLACcitation  = "%%CITATION = HEP-EX 0701033;%%"
% }

\end{footnotesize}

% ****************************************************************************
% END OF BIBLIOGRAPHY AREA
% ****************************************************************************

\end{document}